\documentclass[showpacs,aps,prd,nofootinbib,floatfix,amsmath,amssymb]{revtex4}
\usepackage{graphicx}
\usepackage{multirow}

\begin{document}


\title{Radiative decays $Z_H\to V_i Z$ ($V_i=\gamma, Z$) in little Higgs models}
\author{I. Cort\'es-Maldonado}
\author{A. Fern\'andez-Tellez}
\author{G. Tavares-Velasco}		
\email[E-mail:]{gtv@fcfm.buap.mx}
\affiliation{Facultad de
Ciencias F\'\i sico Matem\'aticas, Benem\'erita Universidad
Aut\'onoma de Puebla, Apartado Postal 1152, Puebla, Pue., M\'
exico}

\date{\today}

\begin{abstract}
The study of the phenomenology of an extra neutral gauge boson, $Z_H$, can help us to
unravel the underlying theory. We study the decay of
such a particle into two neutral gauge bosons, $Z_H\to V_i Z$ ($V_i=\gamma, Z$),
in two popular versions of the little Higgs model: the littlest Higgs model
(LHM) and the simplest little Higgs model (SLHM). These decays are induced at
the one-loop level by a fermion triangle and are interesting as they are
strongly dependent on the mechanism of anomaly cancellation. All
the relevant tree-level two- and three-body decays of the $Z_H$ gauge boson are
also calculated. It is found that the branching ratios for the $Z_H\to \gamma Z$
decays  can be as large as that of a tree-level three-body decay but the
$Z_H\to ZZ$ decay is very suppressed. We also discuss the
experimental
prospects for detecting these decays at the LHC and a future linear collider.
We conclude that the latter would offer more chances for the detection of such
rare
decays.

\end{abstract}

\pacs{14.70.Pw,13.38.Dg}

 \maketitle

\section{Introduction}
\label{int}
The standard model (SM) of electroweak interactions has proven highly successful as its predictions have been confirmed with a high precision at particle colliders. The only missing ingredient of this theory  is the Higgs boson, which plays an essential role in the mechanism of electroweak symmetry breaking (EWSB). A global fit to electroweak precision data collected at LEP and Tevatron  \cite{:2005dia} suggests that the Higgs boson mass, $m_H$, is below 209 GeV. Since $m_H$ receives quadratically divergent contributions at the one-loop level from the top quark, the gauge bosons, and the Higgs boson itself, fine-tuning would be required to get a relatively light $m_H$. The problem would worsen if the new physics scale was of the order of the Planck scale. This is known as the little hierarchy problem, which remains among the unanswered puzzles of the SM. Although supersymmetry has long been known as a promising prospect to solve the hierarchy problem, very recently little Higgs models  \cite{ArkaniHamed:2001nc,ArkaniHamed:2002qy,ArkaniHamed:2002pa,ArkaniHamed:2002qx,Low:2004xc,Kaplan:2003uc,Schmaltz:2004de}  have emerged as an interesting alternative to stabilize the Higgs boson mass without  fine-tuning. This class of theories are based on the old hypothesis that the Higgs boson is a pseudo-Goldstone boson arising from a spontaneously broken
approximate global symmetry at a scale of the order of a few TeVs. After a collective symmetry breaking mechanism is introduced,  there is a set of new particles that play the role of partners of the SM gauge bosons and the top quark. The couplings of these new particles are such that the quadratic divergences to $m_H$ arising at one-loop from the SM particles are exactly
canceled  by the contribution of their respective partners, thereby
yielding a naturally light Higgs boson. Several realizations of this idea have been proposed in the literature, such as the littlest Higgs model (LHM) \cite{ArkaniHamed:2002qy}, the littlest Higgs model with T-parity (LHTM) \cite{Low:2004xc}, which  is actually an extension of the latter, and the simplest little Higgs model (SLHM) \cite{Kaplan:2003uc,Schmaltz:2004de}.

Little Higgs models predict effects that may show up  at the 1 TeV level, so the study of their phenomenology could be at the reach of the large hadron collider (LHC) or a future $e^-e^+$ linear collider. The scale of the global symmetry breaking  as well as other parameters of little Higgs models have been constrained from low energy electroweak measurements
\cite{Csaki:2002qg,Hewett:2002px,Csaki:2003si,Gregoire:2003kr,Chen:2003fm,Casalbuoni:2003ft,Marandella:2005wd,Han:2005dz,
Kilian:2003xt, Hubisz:2005tx} and the respective phenomenology  has been  widely studied throughout the last years
\cite{Burdman:2002ns,Han:2003wu,Han:2005ru,Hubisz:2004ft, Belyaev:2006jh,Freitas:2006vy}.
Even if the new particles predicted by little Higgs models were too heavy to be directly
produced at particle colliders, they could show-up via
loop effects in particle observables. Apart from reproducing the SM at the electroweak scale,  little Higgs models predict heavy partners for the top quark and the weak gauge bosons, which are necessary to cancel the quadratic divergences of the Higgs boson mass at the one-loop level. There can also be a massive partner for the photon, as well as new scalar particles and additional fermions, but their presence is more dependent on the particular implementation of the model. In this work we will concentrate our attention on the extra neutral gauge boson that is the partner of the SM weak gauge boson. As explained below, it can give a more robust signal of the model at particle colliders than a heavy photon.

An extra neutral gauge boson arises in models in which the SM group is extended with an extra  gauge group or if it is embedded into a larger gauge group, such as occurs in the left-right symmetric model, grand unified theories, 331 models, extra dimension theories, technicolor models, the twin-left right symmetric model, etc. It is worth mentioning that the literature has been mainly devoted to the study of the extra neutral gauge boson associated with an extra $U'(1)$ gauge group \cite{Langacker:2008yv}, which is customarily denoted by $Z'$.  In the LHM and its T-parity extension, the $Z$ gauge boson partner, denoted by $Z_H$, is associated with the additional  $SU(2)$ gauge group. In these models there is also a heavy photon partner, $A_H$, which is the lightest new particle and is the analogue of the $U'(1)$ $Z'$ gauge boson.  Although this gauge boson has a great potential to hint the first LHM evidences  at a particle collider, it has been argued  that it would not offer a robust signal of the model  due to the arbitrariness of the charge assignments of the SM fermions under the $U'(1)$ gauge group.  As far as the SLHM is concerned, the role of the $Z$ partner, which is denoted by $Z'$, is played by a linear combination of $SU(3)$ and $U(1)$ gauge fields. This model also predicts a new no self conjugate extra neutral gauge boson, $Y^0$. From now on, the $V$ letter will denote the extra neutral gauge boson that plays the role of the  partner of the $Z$ gauge boson in little Higgs models. When we refer to a particular model version, we will use the customary notation to refer to this extra neutral gauge boson.

It is not possible to obtain a model-independent bound on the mass of an extra neutral
gauge boson from experimental measurements, but electroweak precision data
\cite{Erler:1999nx} along with Tevatron  \cite{:2007sb} and LEP2 \cite{Alcaraz:2006mx}
searches, allow one to obtain limits on $m_{Z'}$ from about $500$ GeV to $1000$ GeV in
models with universal flavor gauge couplings.  Since the mass of the new heavy gauge
bosons predicted by little Higgs models are of the the order of $f$, a bound on $f$
translates into an indirect bound on $m_{V}$. While an extra neutral gauge boson with a
mass around 4-5 TeV may be detected at the LHC,  the future international linear collider
would be able to produce it with a mass up to 2-5 TeV \cite{Langacker:2008yv}. This would
open up potential opportunities to study the phenomenology of this particle and even study
some of its rare decays. Since an extra neutral gauge boson may prove useful to find out
the  particular little Higgs model from which it arises, we are interested in studying its
decay modes into a pair of neutral gauge bosons, $V\to V_i Z$ ($V_i=\gamma, Z$), which
arise at the one-loop level but may have a sizable branching ratio similar to that of a
tree-level three-body decay. These decays, which are interesting as their 
rate is dictated by the mechanism of anomaly cancellation, have
already been studied in the context of a superstring-inspired $E_6$ model
\cite{Chang:1988fq}, the minimal 331 model \cite{Perez:2004jc},
and 5D warped-space models \cite{Perelstein:2010yd}. $Z'$ decays
into three neutral gauge bosons were also studied in the framework of the minimal 331
model \cite{FloresTlalpa:2009kd,FloresTlalpa:2009jh}.

The remainder of this paper is structured as follows. In Section II we present a survey of little Higgs models, with particular emphasis on the gauge sector and the properties of the extra neutral gauge boson $V$ that is the partner of the $Z$ boson. Section III is devoted to present the calculation of the one-loop decays $V\to V_i Z$ ($V_i=Z, \gamma$) in the LHM and the SLHM. In order to calculate the respective branching ratios, we will also discuss the dominant decay modes of the $V$ boson  arising at the tree-level. Finally, Sec. IV will be devoted to discuss the results, including some remarks on the experimental possibilities to measure the $V\to V_i Z$ decays at the LHC and a future linear collider.

\section{The framework of little Higgs models}
\label{model}

The idea that the Higgs boson is light because it is a pseudo-Goldstone boson arising from an approximately broken global symmetry associated with a strongly interacting sector was explored long ago \cite{Kaplan:1983fs,Kaplan:1983sm}. The drawback of those models is that, due to the fact that a Goldstone boson can only have derivative couplings, its gauge and Yukawa couplings would necessarily violate the global symmetry. As a consequence, these interactions would generate radiatively a mass term for the  Goldstone boson, which would be of the same order as the one appearing in models in which no global symmetry is present, thereby preventing a light Higgs boson unless fine-tuning is reintroduced. A solution to this problem was suggested by Arkani-Hamed, Cohen and Georgi \cite{ArkaniHamed:2001nc}. By invoking a collective mechanism of symmetry breaking (the Goldstone bosons are parametrized by a nonlinear sigma model which apart from a global symmetry under the group $G_1$ has a local symmetry under the subgroup $G_2\subset G_1$), the gauge and Yukawa couplings of the Goldstone boson are introduced in such a way that the Higgs boson mass is free of quadratic divergences at the one-loop  or even at the two-loop level. In the fermion sector it  is  necessary
to introduce a new vector-like top quark (top partner) to cancel the quadratically divergent contribution to the Higgs boson mass from the top quark loops.  This idea can be implemented in several ways, but there are basically two different types of little Higgs models \cite{Perelstein:2005ka}: product group models, in which the SM gauge group is the diagonal subgroup  of a larger gauge group, and simple group models, in which the SM gauge group is embedded into a larger gauge group. Models that fall into each category  are the LHM, which is based on the $[SU(2)\times U(1)]^2$ product gauge group, and the SLHM, which has gauge symmetry under the $SU(3)\times U(1)$ simple group. These models share features common to other models of the same class, so  the study of their phenomenology may shed light on the properties of similar models. As  stated above, we will concentrate on two particular little Higgs models, namely, the LHM, \cite{ArkaniHamed:2002qy} and the SLHM \cite{Schmaltz:2004de}. We will not consider the LHM with T-parity as the decays we are interested in are forbidden in that model.  Instead of discussing with detail the theoretical  framework of these models, we will content ourselves with focusing on those topics essential for our discussion. A detailed description can be found in the original works.

\subsection{The littlest Higgs model}

The most economic and most popular version of little Higgs models is the LHM \cite{ArkaniHamed:2002qy}. However, electroweak precision measurements put stringent constraints on the scale of the symmetry breaking, $f$, of the order of 4 TeV \cite{Csaki:2002qg}, rendering the model somewhat unattractive. There is still a small region of parameter space in which $f$ can be as low as 2 TeV. This problem is alleviated if the model is extended by invoking T parity \cite{Low:2004xc}, a discrete symmetry analogue to R-parity, the symmetry introduced in supersymmetric models. However, in this version of the model the $Z_H\to V_i Z$ ($V_i=\gamma, Z$) decays are forbidden due to T-parity.

The LHM  is a  nonlinear sigma model  with a global symmetry under the $SU(5)$ group and a gauged subgroup $[SU(2) \otimes U(1)]^2$. The Goldstone bosons are parametrized by the following $\Sigma$ field
\begin{equation}
\Sigma = e^{i\Pi/f}\ \Sigma_0\  e^{i\Pi^T/f}
\label{sigmaA}
\end{equation}
where $\Pi$ is the pion matrix. The $\Sigma$ field transforms under the gauge group as $\Sigma \to \Sigma' = U\  \Sigma \ U^T$,
with  $U=L_1 Y_1 L_2 Y_2$ an element of the gauge group.

The $SU(5)$ global symmetry is broken down to
$SO(5)$ by the sigma field VEV, $\Sigma_0$, which is of the order of the scale of the symmetry breaking. After the global symmetry is broken, 14 Goldstone bosons arise accommodated in multiplets of the electroweak gauge group: a real singlet, a real triplet, a complex triplet and a complex doublet. The latter will be identified with the SM Higgs doublet. At this stage, the gauge symmetry is also broken down to its diagonal subgroup,  $SU(2) \times U(1)$.  The real singlet and the real triplet are absorbed by the gauge bosons associated with the broken gauge symmetry.

The LHM effective Lagrangian is assembled by the kinetic
energy Lagrangian of the $\Sigma$ field, $\mathcal{L}_{\rm K}$,
the Yukawa Lagrangian, $\mathcal{L}_{\rm Y}$, and the kinetic terms of the gauge and fermion sectors.  The sigma field kinetic Lagrangian is given by
\cite{ArkaniHamed:2002qy}
\begin{equation}
\mathcal{L}_{\rm K}= \frac{f^{2}}{8} {\rm Tr} | D_{\mu} \Sigma |^2,
\label{kinlag}
\end{equation}
with the $[SU(2)\times U(1)]^2$ covariant derivative defined by \cite{ArkaniHamed:2002qy}
\begin{equation}
D_{\mu} \Sigma = \partial_{\mu} \Sigma
- i \sum_{j=1}^2 \left[ g_{j} W_{j\,\mu}^{a} (Q_{j}^{a}\Sigma + \Sigma Q_{j}^{a\,T})
+ g'_{j} B_{j\,\mu} (Y_{j} \Sigma+\Sigma Y_{j}^{T}) \right].
\end{equation}
The heavy $SU(2)$ and $U(1)$ gauge bosons are  $W_{j}^\mu =
\sum_{a=1}^{3} W_{j}^{\mu \, a} Q_{j}^{a}$ and $B_{j}^\mu =
B_{j}^{\mu} Y_{j}$, with  $Q_j^a$ and $Y_j$ the gauge generators,
while $g_i$ and $g'_i$ are the respective gauge couplings. The VEV
$\Sigma_0$ generates masses for the gauge bosons and mixing between them. The heavy gauge boson mass eigenstates are given
by \cite{ArkaniHamed:2002qy}
\begin{eqnarray}
W'^a &=& -c W_{1}^a + s W_{2}^a,\\
B' &=& -c^{\prime} B_{1} + s' B_{2}  ,
\end{eqnarray}
with masses  $m_{W'} = \frac{f}{2}\sqrt{g_1^2+g_2^2}$ and
$m_{B'} = \frac{f}{\sqrt{20}} \sqrt{g'^2_{1}+g'^2_{2}}$.

The orthogonal combinations of gauge bosons are identified with the SM
gauge bosons:
\begin{eqnarray}
W^a &=& s W_{1}^a + c W_{2}^a,\\
B &=& s' B_{1} + c' B_{2},
\end{eqnarray}
which remain massless at this stage,
their couplings being given by
$g = g_{1}s = g_{2}c$ and
$g' = g'_{1}s' = g'_{2}c'$,
where $s = g_{2}/\sqrt{g_{1}^{2}+g_{2}^{2}}$ and
$s' = g'_{2}/\sqrt{g'^2_{1}+g'^2_{2}}$ are mixing parameters (here $c=\sqrt{1-s^{2}}$ and $c' = \sqrt{1-s'^{2}}$).

The gauge and Yukawa interactions that break the global $SO(5)$ symmetry induce radiatively a Coleman-Weinberg potential, $V_{CW}$, whose explicit form can be obtained after expanding the $\Sigma$ field:
\begin{equation}
    V_{\rm CW} = \lambda_{\phi^2} f^2 {\rm Tr}|\phi|^2
    + i \lambda_{h \phi h} f \left( h \phi^\dagger h^T
        - h^* \phi h^\dagger \right)
    - \mu^2 |h|^2
    + \lambda_{h^4}  |h|^4,
\end{equation}
where $\lambda_{\phi^2}$, $\lambda_{h \phi h}$, and $\lambda_{h^4}$
depend on the fundamental parameters of the model, whereas $\mu^2$, which receives logarithmic divergent
contributions at one-loop level and quadratically divergent
contributions at the two-loop level, is treated as a free parameter
of the order of $f^2/16 \pi^2$. The Coleman-Weinberg potential induces a mass term for the complex triplet $\Phi$, whose components acquire a mass of the order of $f$. The neutral component of the complex doublet develops a VEV, $v$, of the order of the electroweak scale, which is responsible for EWSB. The VEV $v$ along with the triplet VEV, $v'$, are obtained when $V_{CW}$ is minimized.

At the electroweak scale, EWSB proceeds as usual, yielding the final mass eigenstates: the three SM gauge bosons are accompanied by three heavy gauge bosons which are their counterpart,  $A_H$, $W_H$ and
$Z_H$. The masses of the heavy gauge bosons  get
corrected by terms of the order of $(v/f)^2$ and so are the masses
of the weak gauge bosons $W_L$ and $Z_L$. The heavy gauge boson masses are given by
\cite{Han:2003wu}:

\begin{equation}
m_{Z_H}^2\simeq m_{W_H}^2=m_W^2\left(\frac{f^2}{s^2c^2v^2}-1\right)\ge 4 m_W^2 \frac{f^2}{v^2}, \label{WHmasslimit}\\
\end{equation}
\begin{equation}
m_{A_H}^2= m_Z^2 s_W^2\left(\frac{f^2}{5 s'^2c'^2v^2}-1+\frac{x_H c_W^2}{4s^2 c^2 s_W^2}\right)\ge 4 m_W^2 t_W^2 \frac{f^2}{5v^2}, \label{AHmasslimit}
\end{equation}
with $t_W=s_W/c_W$, being $s_W$ and $c_W$ the sine and cosine of the Weinberg angle $\theta_W$, while $x_H=\frac{5}{2}gg'\frac{scs'c'(c^2s'^2+s^2c'^2)}{5 g^2s'^2c'^2-g'^2s^2c^2}$.

In the scalar sector, after diagonalizing the Higgs mass matrix, the light Higgs boson
mass can be obtained at the leading order \cite{Han:2003wu}
\begin{equation}
    m^2_{H}= 2 \mu^2 = 2 \left( \lambda_{h^4}
    - \frac{\lambda_{h \phi h}^2}{ \lambda_{\phi^2}} \right) v^2
\end{equation}

It is required that  $\lambda_{h^4} > \lambda_{h \phi h}^2 /
\lambda_{\phi^2}$ to obtain the correct  electroweak symmetry
breaking vacuum with $m^2_H>0$. The Higgs triplet masses are
degenerate at this order:

\begin{equation}
m_{\Phi}=\sqrt{2} m_H\frac{f}{v},
\label{Phimass}
\end{equation}

In summary, in the gauge sector there are four new gauge bosons $W_H^\pm$, $Z_H$
and $A_H$, while in the scalar sector there are new neutral, singly charged and doubly charged Higgs
scalars, $\phi^0$, $\phi^-$, $\phi^{--}$, together with one
pseudoscalar boson $\phi^P$. The presence of the heavy gauge bosons $W_H$ and $Z_H$ is generic in little
Higgs models since they are necessary for the collective symmetry mechanism. However, the scalar sector depends on the particular implementation of the model.

\subsubsection{Fermion sector}

The fermion sector  is identical to the SM one except in the top sector, which requires a new
vector-like top quark $T$, which is known as the top partner. The $T$ loops
cancel the quadratically divergent contribution to the Higgs mass arising
from the top quark loops. This fixes the Yukawa interactions, given
by \cite{ArkaniHamed:2002qy}

\begin{equation}
\mathcal{L}_{\rm Y} = \frac{1}{2}
\lambda_{1} f \epsilon_{ijk} \epsilon_{xy} \chi_{i}
\Sigma_{jx}\Sigma_{ky} {u'}_{3}^{c}
+ \lambda_{2} f \tilde{t} \tilde{t}^{c} + {\rm H.c.},
\end{equation}
where $\epsilon_{ijk}$ and $\epsilon_{xy}$ are antisymmetric tensors. The subscripts $i,j$ ($x,y$) are summed over $1..3$ ($4..5$). In addition, $t_3$ is the SM top quark, $u'_{3}$ is the SM right-handed top
quark, $({\tilde t},\tilde{t'}^{c})$ is a new vector-like top quark
and ${\chi}=(b_{3}, t_{3},  \tilde{t})$. The first term of
${\mathcal L}_{\rm Y}$ induces the couplings of the Higgs boson to
the fermions such that the quadratic
divergences from the top quark loop are canceled by the top partner loop.
The expansion of the $\Sigma$ field leads to the physical states,
$t$ and $T$, after diagonalizing the mass matrix. At the leading
order in $v/f$, the masses of the SM top quark and the new top quark $T$ are given by
\cite{Han:2003wu}

\begin{equation}
    m_t = \frac{\lambda_1 \lambda_2}{\sqrt{\lambda_1^2 + \lambda_2^2}} v,\quad
    m_T = f \sqrt{\lambda_1^2 + \lambda_2^2}.\label{Topmass}
\end{equation}

There is no need to introduce extra vector-like quarks for the first two quark generations as the quadratic divergences arising from light fermions are not important below the cutoff scale $\Lambda_S=4\pi f$.

The remaining terms of the LHM Lagrangian and all the Feynman rules for the new
interactions were given  in \cite{Burdman:2002ns,Han:2003wu}.  In particular, the
couplings of the heavy neutral gauge bosons depend on the isospin and hypercharge of the
fermions, which is dictated by the gauge invariance of the scalar couplings to the
fermions under $U(1)_1\times U(1)_2$. They are given by
\begin{equation}
{\cal L}=\frac{g'}{s'c'}\left(-{c'}^2 J^\mu_{B_1}+{s'}^2
J^\mu_{B_2}\right){A_H}_\mu+\frac{gc}{s}J^\mu_{W^3} {Z_{H}}_\mu+ {\rm H.c.},
\label{ZHffLHM}
\end{equation}
with  $J^\mu_{W^3}=\bar{Q}_L\gamma^\mu(T^3)Q_L$ and $J^\mu_{B_{1,2}}\bar{f}\gamma^\mu Y_{1,2}f$, while $Y_{1,2}$ represent the $U(1)_{i,j}$ quantum number assignments of the $\Sigma$ field. The fermion hypercharges are given in terms of two free parameters,  $y_u$ and $y_e$, which can be fixed to $y_u=2/5$ and $y_e=3/5$ by requiring anomaly cancellation under both $U(1)$ groups.

The Feynman rules for all the couplings necessary for the calculation of the   decays of
the $Z_H$ gauge boson were taken from \cite{Burdman:2002ns,Han:2003wu} and are presented
in Appendix \ref{Couplings} for completeness.

\subsection{The simplest little Higgs model}

This realization of little Higgs models  has a global symmetry under the group $[SU(3)\times U(1)]^2$ and a gauged subgroup $SU(3)\times U(1)$ \cite{Schmaltz:2004de}. It is necessary to introduce  two sigma fields $\Phi_1$ and $\Phi_2$, whose  VEVs  break down the global symmetry down to the subgroup $[SU(2)\times U(1)]^2$, generating 10 Goldstone bosons.  At this stage, the sigma field VEVs also break the gauged subgroup down to the SM gauge group. The kinetic Lagrangian of the sigma model can be written as

\begin{equation}
 {\cal L}=\sum_{i=1,2}|D_\mu \Phi_i|^2,
\label{kinslh}
\end{equation}
with the sigma fields given by
\begin{equation}
\Phi_1 = e^{i\Theta_1/f_1}<\Phi_{(3,1)}>
\label{sigmaphi1}
\end{equation}
and
\begin{equation}
\Phi_1 = e^{i\Theta_2/f_2}<\Phi_{(1,3)}>
\label{sigmaphi2}
\end{equation}
where the VEVs of the sigma fields, $<\Phi_{(3,1)}>$ and $<\Phi_{(1,3)}>$, are of the order of $f_1\sim f_2\sim 1$ TeV. Here the subscripts denote the VEVs transformation properties under the $SU(3)$ group. Also, $\Theta_{1,2}$ stand for the pion matrices. The covariant derivative of the $SU(3)\times U(1)$ can be written as

\begin{equation}
D_\mu=\partial_\mu-i gA_\mu^a T^a-i\frac{g_X}{3}B_\mu,
\end{equation}
where $A_\mu^a$ ($i=1 .. 8$) and $B_\mu$ are the gauge fields of the $SU(3)$ and $U(1)$ gauge groups, $T^a$ stands for the $SU(3)$ generators, whereas $g$ and $g_X$ are the associated gauge coupling constants. The latter is required to be $g_X=\sqrt{3}gt_W/\sqrt{3-t_W^2}$ to match the SM hypercharge coupling constant. After the breaking of the local symmetry, five massive gauge bosons emerge in a complex doublet $(X^{\pm},Y^0)$ and a real singlet $Z'$ of $SU(2)_L$, which are given in terms of the gauge fields as:

\begin{eqnarray}
Y^{0} &=&\frac{1}{\sqrt{2}}\left(A^4 \mp iA^5\right), \\
X^\pm&=&\frac{1}{\sqrt{2}}\left(A^6 \mp iA^7\right),\\
\end{eqnarray}
The extra neutral gauge boson $Z'$ is a linear combination of $A^8$ and $B^X$:

\begin{equation}
 Z'=\frac{\sqrt{3}gA^8+g_XB^X}{\sqrt{3g^2+g_X^2}}.
\end{equation}
Five of the ten generated Goldstone bosons are eaten by the heavy gauge bosons, which get masses of the order of $f=\sqrt{f_1^2+f_2^2}$, whereas the remaining ones accommodate in a real singlet $\eta$ and a complex doublet $h$ of the electroweak gauge group. We can identify the electroweak gauge fields as follows

\begin{eqnarray}
W^3&=&A^3,\\
W^\pm&=&\frac{1}{\sqrt{2}}\left(A^1 \mp i A^2\right),\\
B&=&\frac{-g_X A^8+\sqrt{3}B^X}{\sqrt{3g^2+g_X^2}}.
\end{eqnarray}

The scalar complex doublet $h$ corresponds to the SM Higgs doublet and develops a VEV, via a radiatively generated Coleman-Weinberg potential induced by the gauge and Yukawa interactions. EWSB is triggered as usual, after which the SM gauge bosons acquire mass and the heavy gauge bosons get additional mass terms. The heavy and light physical states as well as their masses can be obtained after expanding Eq. (\ref{kinslh}) in powers of $v/f$.
Up to order $(v/f)^2$, the Lagrangian for the charged gauge bosons $W^\pm$ and $X^{\pm}$ is diagonal and their masses are
\begin{equation}
 m_W=\frac{gv}{2},
\end{equation}
\begin{equation}
 m_{X}=\frac{gf}{\sqrt{2}}\left(1-\frac{v^2}{4f^2}\right),
\end{equation}
The charged physical states differ from the gauge eigenstates by terms of the order of  $(v/f)^3$. Thus, unless a high precision is required, the charged gauge eigenstates can be considered the same as the mass eigenstates. As far as the neutral gauge bosons are concerned, the gauge eigenstates must  be rotated to obtain the physical eigenstates at the order $(v/f)^2$.  The $Z'$ gauge boson gets mixed with the SM $Z$ boson by a term of the order of $(v/f)^2$. The  physical states are obtained after the replacement:
$Z'\to Z' +\delta_Z Z$ and $Z\to Z-\delta_Z Z'$, with
\begin{equation}
\delta_Z=-\frac{(1-t_W^2)\sqrt{3-t_W^2}}{8c_W}\frac{v^2}{f^2}.
\end{equation}
The masses of the massive neutral gauge bosons are given by
\begin{equation}
 m_{Z}=\frac{gv}{2c_W},
\end{equation}
\begin{equation}
 m_{Y}=\frac{gf}{\sqrt{2}}\left(1-\frac{v^2}{4f^2}\right),
\end{equation}
\begin{equation}
 m_{Z'}=\frac{\sqrt{2}gf}{\sqrt{3-t_W^2}},
\end{equation}
In summary, apart from the SM gauge spectrum, in the gauge sector there are a pair of new heavy charged gauge bosons, $X^{\pm}$, a new no self-conjugate neutral gauge boson, $Y^0$, and a new neutral gauge boson, $Z'$. The latter plays the role of the SM $Z$ gauge boson and is the focus of this paper. As far as the scalar sector is concerned, there is a new neutral  scalar, which can be light but has a different phenomenology than the one of the SM Higgs boson.

\subsubsection{Fermion sector}
In the fermion sector, the $SU(2)_L$ doublets need to be promoted to $SU(3)_L$ triplets, which requires the inclusion of new fermions  along with new right singlets to endow the new fermions with masses. Notice that while the inclusion of the top partner is necessary to cancel the  Higgs boson mass quadratic divergences at the one-loop level, it is not necessary to include additional partners for the light fermions.   The three lepton families transform similarly under the gauge group and include one new neutral lepton $N_i$ for each generation:
\begin{equation}
{ l _i}_L=\left( \begin{array}{ccc} \nu_i \\
e_i\\
iN_i
\end{array}\right) \sim (3,-1/3), \ \  i{e_i}^c \sim (1,0),\ \  i{N_i}^c \sim
(1,-1),
\end{equation}
where $i=1,2,3$ stands for the family index and the gauge quantum numbers appear
in the parenthesis. In the quark sector, a new quark for each family is
necessary. Two different alternatives have been proposed to add the new quarks:
the universal embedding and the anomaly-free embedding.

In the universal embedding the three quark generations carry identical $SU(3)_L$
quantum numbers. They transform as
\begin{equation}
{Q_i}_L=\left( \begin{array}{ccc} {u_i} \\
{d_i}\\
i{U_i}
\end{array}\right) \sim (3,1/3), \ \
i{u_i}^c \sim (1,-2/3),  \ \  i{d_i}^c \sim (1,1/3),\, \ i{U_i}^c
\sim (1,-2/3),
\end{equation}
for $i=1,2,3$. The three new quarks are $U_1=U$, $U_2=C$, and $U_3=T$, which are
partners of the $u$, $c$ and $t$ quarks, respectively.
This leads to $SU(3)_L\times U(1)_X$ anomalies although the $SU(2)_L\times
U(1)_Y$ gauge group remains anomaly free. Since the SLHM is an effective theory
valid up to the cut-off scale $\Lambda_S$, the anomalies must be canceled by new
fermions included in the ultraviolet completion of the theory.

Another alternative is to choose a particular transformation for the triplets
such that anomalies are canceled \cite{Kong:2003tf}. In this case each quark
generation has different quantum number assignments. While the first two
families transform alike:
\begin{equation}
{Q_{1,2}}_L=\left( \begin{array}{ccc} {d_{1,2}} \\
-{u_{1,2}}\\
i{D_{1,2}}
\end{array}\right) \sim (\bar{3},0), \ \
i{d_{1,2}}^c \sim (1,1/3),  \ \  i{u_{1,2}}^c \sim (1,-2/3), \ \ i{D_{1,2}}^c
\sim (1,1/3),
\end{equation}
the third family transforms differently
\begin{equation}
{Q_3}_L=\left( \begin{array}{ccc} {b} \\
{t}\\
T_L
\end{array}\right) \sim (3,1/3),\ \
{b}^c\sim (1,1/3),\ \ i{t}^c\sim (1,-2/3),\ \ T^c\sim
(1,-2/3).
\end{equation}
The three new quarks are $D_1=D$, $D_2=S$, and $T$, which  are partners of the
SM quarks $d$, $s$ and $t$, respectively. It is worth mentioning that anomaly
cancellation does not occur family by family as in the SM but only when the
three families are summed over. This mechanism of anomaly cancellation is
identical to that introduced in the 331 model with right-handed neutrinos
\cite{Foot:1994ym}.

.

The Yukawa Lagrangian for both the lepton and the quark sector along with the
Lagrangians for the gauge sector and the fermion sector were worked out with
detail in \cite{Han:2005ru}. The fermion masses are given by:

\begin{eqnarray}
m_{N_i}&=&\lambda_{N_{i}} s_\beta f,\\
m_{Q_i}&=&\lambda_{Q_{i}} s_\beta f,\\
m_T&=&\sqrt{\lambda_1^2 c_\beta^2+\lambda_2^2 s_\beta^2}f,\\
\end{eqnarray}
where $\tan\beta=f_1/f_2$, and the usual notation, $s_\beta=\sin\beta$ and
$c_\beta=\cos\beta$, has been introduced.

The Feynman rules necessary for the calculation of the $Z'$ decays are taken
from Ref. \cite{Han:2005ru} and appear in  Appendix \ref{Couplings}.

\section{Extra neutral gauge boson decays}
\label{cal}
We now present the analytical results for the calculation of the extra neutral
gauge boson decays in the models discussed above. We begin with the tree-level
decays and afterwards focus on the one-loop induced $V\to V_iZ $ ($V_i=\gamma,
Z$) decays.
\subsection{Littlest Higgs model}
\subsubsection{Tree-level two-body and three-body decays}
The dominant decays of the neutral gauge boson $Z_H$ are the tree-level induced
two-body decays $Z_H\to \bar{f}f$, $Z_H\to W^+W^-$, $Z_H\to ZH$, and $Z_H\to A_H
H$. The latter is the only  kinematically allowed tree-level two-body decay
involving a new particle as a final state. The calculation is straightforward
and we will present  the respective decay widths in a rather generic form, which
will be useful for  the SLHM calculations. We first present the decay width into
the fermion pair $\bar{f}f$ assuming an interaction similar to that given in Eq.
(\ref{ZHffLag}) for the coupling of the extra neutral gauge boson to a fermion
pair:

\begin{equation}
\Gamma(V \rightarrow f\bar{f}) =\frac{g^2m_{V} N_c^f}{24
   \pi c_W^2} \sqrt{1-4 y_f} \left(\left({g'_L}^2+{g'_R}^2\right)(1-y_f)
   +6 {g'_L}
   {g'_R}y_f\right)
\label{Vtoff},
\end{equation}
were $V$ represents the extra neutral gauge boson and we introduced the notation
$y_a=(m_a/m_{V})^2$. $N_c^f$ is the fermion color number.

The $V\to W^+W^-$ and $V\to ZH$ decay widths are

\begin{equation}
\Gamma(V \rightarrow WW) = \frac{g_{VWW}^2 m_{V}}{192 \pi
   y_W^2} \left(1-4 y_W\right)^{3/2} \left(1+
   20 y_W+12y_W^2\right),
\label{VtoWW}
\end{equation}

\begin{equation}
\Gamma(V \rightarrow ZH) \frac{g_{VZH}^2} {192 \pi m_{V}y_Z}
\sqrt{(1-(\sqrt{y_H}-\sqrt{y_Z})^2)(1-(\sqrt{y_H}+\sqrt{y_Z})^2)}
\left(1+(y_H-y_Z)^2+y_Z^2-2(y_H-5y_Z)\right).
\label{VtoZH}
\end{equation}
The $Z_H\to A_H H$ decay width can be obtained from $\Gamma(Z_H\to ZH)$ after
the replacements $y_Z\to y_{A_H}$ and $g_{Z_H ZH}\to g_{Z_H A_H H}$ are done. The above
results agree with the calculation presented in \cite{Burdman:2002ns}.
We also calculated the tree-level three-body decays into SM particles: $Z_H\to
\bar{f}f\gamma$, $Z_H\to \bar{f}fZ$, $Z_H\to \bar{t}tH$, $Z_H\to ZHH$, $Z_H\to
ZW^-W^+$, $Z_H\to
\gamma W^-W^+$, and $Z_H\to ZZZ$. The latter is mediated by a virtual Higgs
boson. For completeness, we also calculate other kinematically allowed
three-body decays involving a heavy photon:  $Z_H\to A_H H H$, $Z_H\to A_H W W$,
$Z_H\to A_H Z Z$, and $Z_H\to A_H A_H A_H$. To obtain the decay widths,
we squared the decay amplitude with the aid of the FeynCalc package and the
integration over the three-body phase space was performed numerically. We
refrain from presenting the analytical results as they are too cumbersome to be
included here. It is worth noting that the two- and three-body decays of the
$Z_H$ gauge boson into SM particles were already studied in \cite{Boersma:2007fd} for
values of the mixing angle $c$ such that  the
couplings of the $Z_H$ gauge boson to scalar Higgs bosons become strongly
interacting. We also note that the decays widths for $Z_H\to
\bar{f}f\gamma$ and $Z_H\to \gamma W^-W^+$ were obtained with the assumption of
$E_\gamma\ge 10$ GeV to avoid infrared divergences. All the coupling constants
appearing above can be found in Appendix \ref{Couplings}.

\subsubsection{One-loop decays $Z_H\to V_iZ $ ($V_i=\gamma,Z$)}
\begin{figure}
 \centering
\includegraphics*[width=4.5in]{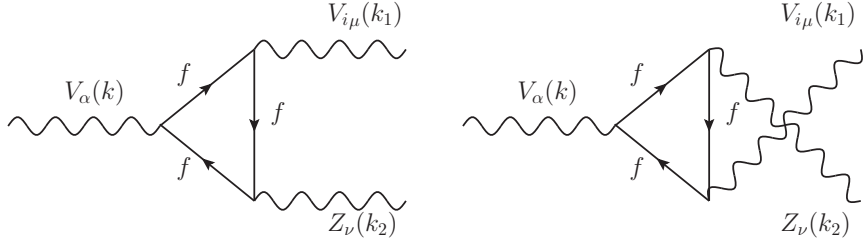}
 \caption{\label{FeynDiag1}One-loop Feynman diagrams contributing to the extra
neutral gauge boson decay
 $V\to V_iZ$, with $V_i=\gamma, Z$ in the LHM and the SLHM. This decay is
forbidden in the LHTM.}
 \end{figure}

We now turn to  the one-loop level  two-body decays $Z_H\to
\gamma Z$ and $Z_H \to ZZ$. It is worth mentioning that the $Z_H$ decay into a
photon pair is forbidden by the Landau-Yang theorem. We will not consider the
one-loop decays involving a heavy photon as they are expected to have a smaller
decay width due to phase space suppression. In the LHM the decay $Z_H \to V_iZ$ 
is induced by the fermion triangle shown in Fig. \ref{FeynDiag1}. The same
fermion circulates through the loop as we will not consider those Feynman
diagrams induced by the nondiagonal vertices $Z_H \bar{T} t$ and $Z \bar{T}t$
(the corresponding amplitude is suppressed by powers of $v/f$). Also, the
charged gauge boson loops  do not contribute  
to trilinear neutral gauge boson vertices. This can be explained from the fact
that these kind of contributions cannot  generate the
structure of Eq. (\ref{MVtoAZ}), which involves the Levi-Civitta tensor. The
decay amplitudes were calculated via the
Passarino-Veltman technique \cite{Passarino:1978jh} via the FeynCalc package
\cite{Mertig:1990an}. After the mass-shell and transversality conditions for the
gauge bosons were considered, and once we got rid of superfluous terms via the
the Schouten identity, the $V \to \gamma Z$ decay amplitude can be cast in the
form

\begin{equation}
{\cal M}(V\to \gamma Z)=\frac{i}{m_{V}^2}\left(A_1^{\gamma Z} \left(k_1^\nu
\epsilon^{\alpha\mu\lambda\rho}+{k_1}^\alpha \epsilon^{\mu \nu
\lambda\rho}\right){k_1}_\lambda {k_2}_\rho+A_2^{\gamma Z}k_1\cdot k_2 \,
\epsilon^{\alpha\mu\nu\lambda}{k_1}_\lambda\right)
\epsilon_{\alpha}(k)\epsilon_{\mu}(k_1)\epsilon_{\nu}(k_2),
\label{MVtoAZ}
\end{equation}
where the four-momenta ${k_1}_\mu$ and ${k_2}_\nu$ correspond to the outgoing
$\gamma$ and $Z$ gauge bosons, respectively. The mass factor and the scalar
products were included for convenience purpose only. The above amplitude
displays explicitly electromagnetic gauge invariance. Assuming  the couplings
given in Eq. (\ref{ZHffLag}), the $A_i^{\gamma Z}$  coefficients can be written
in terms of Passarino-Veltman scalar functions as follows

\begin{eqnarray}
A_1^{\gamma Z}&=&\left(\frac{g}{8\pi c_W}\right)^2 \frac{2}{\left(1 -
y_{Z}\right)^3}\sum_{f}\xi^f_{\gamma Z}  \Big(y_Z^2\left(2+B_c-B_a+2(2 y_f+  1
)C_a\right)+ B_a-B_c\nonumber\\&-&2y_Z\left(1+3 (B_b- B_c)+(2y_f+
1)C_a\right)\Big),
\label{A1VtoAZ}
\end{eqnarray}
\begin{eqnarray}
A_2^{\gamma Z}&=&\left(\frac{g}{8\pi c_W}\right)^2 \frac{2}{\left(1 -
y_{Z}\right)^3}\sum_f\Bigg( \xi^f_{\gamma Z}\Big(y_Z^2 \left(B_a-B_c-2(1+ C_a)
\right)+B_c-B_a-4 y_f  C_a\nonumber\\&+&2
y_Z\left(1+B_b-B_c+(2y_f+1)C_a\right)\Big)+4\lambda^f_{\gamma Z}(1-y_Z)^2y_f
C_a\Bigg),
\label{A2VtoAZ}
\end{eqnarray}
where $\xi^f_{\gamma Z}= N_c^f Q^f  \left({g'_L}^f {g_L}^f-{g'_R}^f
{g_R}^f\right)$ and $\lambda^f_{\gamma Z}=N_c^f Q^f  \left({g'_L}^f
{g_R}^f-{g'_R}^f {g_L}^f\right)$, with $Q_f$ the fermion electric charge,
$N_c^f$ the fermion color number and $g_{L,R}^f$ the couplings of the $Z$ gauge
boson to the fermions.  The sum is over all the charged fermions. Anomaly
cancellation requires that $\sum_f \xi^f_{\gamma Z}=0$. $B_i$ stands for the
following two-point scalar functions $B_a=B_0(0, m_f^2, m_f^2)$,
$B_b=B_0(m_{V}^2, m_f^2, m_f^2)$, and $B_c= B_0(m_Z^2, m_f^2, m_f^2)$, while
$C_a=m_{V}^2C_0(0, m_Z^2, m_{V}^2, m_f^2, m_f^2, m_f^2)$ is a three-point scalar
function scaled by the $m_V$ mass. It is evident that the $A_i^{\gamma Z}$
coefficients are free of ultraviolet divergences. After the above amplitude is
squared and  summed (averaged) over polarizations of outgoing (ingoing)
particles, we obtain the following  decay width
\begin{equation}
\Gamma(V\to \gamma Z)=\frac{1}{3}\frac{\left(1-y_Z\right)^5 \left(1+y_Z\right)
m_{V}}{2^5 \pi y_Z}
 |A_1^{\gamma Z}-A_2^{\gamma Z}|^2.
\label{VtoAZ}
\end{equation}

We now concentrate on the $Z_H\to ZZ$ decay. The respective amplitude must obey
Bose symmetry and can be written as

\begin{equation}
{\cal M}(V\to ZZ)=\frac{iA^{ZZ}}{m_{V}^2}\left(k_1^\nu
\epsilon^{\alpha\mu\lambda\rho}+k_2^\mu \epsilon^{\alpha\nu
\lambda\rho}\right){k_1}_\lambda
{k_2}_\rho\epsilon_{\alpha}(k)\epsilon_{\mu}(k_1)\epsilon_{\nu}(k_2),
\label{MVtoZZ}
\end{equation}
where ${k_1}_\mu$ and ${k_2}_\nu$ are the four-momenta of the outgoing $Z$ gauge
bosons. The coefficient $A^{Z Z}$ is

\begin{eqnarray}
A^{ZZ}&=&\frac{g^3}{8\pi^2 c_W^3}\frac{1}{(1 - 4y_Z)^2}\sum_f\Bigg( \big(1+
4 y_Z^2 \left(B_b-B_c+ (1+2y_f)C_b+2\right)-2 y_Z \left(3-(1-4y_f)(B_b- B_c)+
5y_f C_b\right)\nonumber\\&+&2y_f \left((B_b-Bc)+ C_b\right)-4y_Z^3 C_b
\big)\xi^f_{Z Z}+2 y_f \left(1-4 y_Z\right)\left(B_b-B_c+y_Z
C_b\right)\lambda^f_{Z Z}\nonumber\\&-&2  y_f\left(1-4 y_Z\right) \left(2
(B_b-B_c)+ \left(1-2
y_Z\right)C_b\right)\rho^f_{ZZ}\Bigg),
\label{AiVtoZZ}
\end{eqnarray}
with $\xi^f_{Z Z}=N_c^f({g'_L}^f {g^f_L}^2 - {g'_R}^f {g^f_R}^2)$,
$\lambda^f_{ZZ}=N_c^f({g'_L}^f {g^f_R}^2 - {g'_R}^f {g^f_L}^2)$, and
$\rho^f_{ZZ}= N_c^fg_L^f g_R^f ({g'_L}^f-{g'_R}^f)$. The two-point scalar
functions $B_b$ and $B_c$ were given above, while the scaled three-point scalar
function is $C_b=m_{V}^2C_0(m_Z^2, m_Z^2, m_{V}^2, m_f^2, m_f^2, m_f^2)$. The
sum is now over all fermions. Anomaly cancellation requires that $\sum_f\xi^f_{Z
Z}=0$. The decay width is given by

\begin{equation}
\Gamma(V\to ZZ)=\frac{1}{3}\frac{m_{V}}{2^7 \pi y_Z}(1 - 4y_Z)^{5/2}|A^{ZZ}|^2.
\label{VtoZZ}
\end{equation}

We will examine below the behavior of the branching ratios of all the above
decays as functions of the symmetry breaking scale $f$ and the mixing angle $c$.
The results will be discussed in Sec. IV.

\subsection{Simplest Littlest Higgs model}

In this model the main tree-level decays allowed by kinematics are  $Z'\to
\bar{f}f$, $Z'\to W^+W^-$, $Z'\to
ZH$, $Z'\to \bar{f}{f}\gamma$, $Z'\to \bar{f}{f}Z$, $Z'\to \bar{t}tH$,
$Z'\to ZHH$, $Z'\to ZW^-W^+$, $Z'\to AW^-W^+$, and $Z'\to ZZZ$. There are no
decays into the heavy gauge bosons $X$ nor $Y$. In the case of the two-body
decays, we can readily use the expressions given above, Eqs.
(\ref{Vtoff})-(\ref{VtoZH}), after replacing the respective coupling constants
and the extra neutral gauge boson mass. The same method described above was used
for the calculation of the three-body decays. As far as the one-loop two-body
decays $Z'\to \gamma Z$ and $Z'\to ZZ$ are concerned, Eqs. (\ref{MVtoAZ})
-(\ref{VtoZZ}) are also valid. We only need to insert the proper coupling
constants of the $Z'$ gauge boson to a fermion pair. We have calculated the
branching ratios for all these decays  as functions of the scale $f$. The
results are shown in Sec. IV.

\section{Numerical discussion results and experimental perspectives}

We now turn to present the numerical results for the branching ratios of the
extra neutral gauge boson decays in the previously discussed  versions of the
little Higgs model. We will first discuss the current constraints on the
symmetry breaking scale $f$.

In the original LHM, the bounds on the scale $f$ depend on  the mixing angles of
the gauge sector: $\tan \theta=s/c=g_1/g_2$ and $\tan \theta'=s'/c'=g'_1/g'_2$.
Such constraints can be obtained from electroweak precision measurements, such
as the $Z$ pole data,
low-energy neutrino-nucleon scattering, and measurements of the $W$ mass
\cite{Csaki:2002qg,Hewett:2002px,Csaki:2003si,Gregoire:2003kr,Chen:2003fm,
Casalbuoni:2003ft,Marandella:2005wd,Han:2005dz,
Kilian:2003xt}.  The largest
corrections to electroweak precision observables  arise from the heavy gauge
bosons
\cite{Csaki:2002qg}.  A global fit  to experimental
data severely constrains the  symmetry breaking
scale, $f>4$ TeV, for a wide region of values of the mixing parameters
\cite{Csaki:2002qg}. This would require reintroducing  fine-tuning to have a
light Higgs boson, which would render the model somewhat unattractive. It was
suggested however that dangerous corrections to electroweak precision
observables could be
controlled by tuning the parameters of the model \cite{Han:2003wu}, which would
allow
for a less stringent constraint on the scale $f$. In fact, if a suitable
assignment of the quantum numbers of the light fermions under the new $U(1)$
gauge group is chosen, $f$ can be as low as 1-2 TeV in a small region of the
parameter space  \cite{Csaki:2002qg}.  Another solution requires the
introduction of T-parity into the model \cite{Low:2004xc}, which forbids a
triplet VEV $v'$ and cancels the tree-level contributions to electroweak
observables arising from the heavy gauge bosons. In this scenario,  the
constraint on the scale $f$ is
significantly weaker than in the original LHM: $f$ can be as
low as 500 GeV \cite{Hubisz:2005tx}. For the latest direct bounds on the LHM with
T-parity see also \cite{Perelstein:2011ds}. However, in this model the
$Z_H$ gauge boson can only decay into a heavy photon plus additional SM
particles and so the decays we are interested in are forbidden. As far as the
SLHM is concerned, it was argued that the addition  of an explicit quartic Higgs
coupling  yields a region of the parameter space in which the constraints from
electroweak precision measurements are naturally satisfied, thereby allowing for
a less restrictive bound on $f$. However, the actual constraint on $f$ is of the
order of $4$ TeV as shown in \cite{Csaki:2003si}, which does not necessarily
spoils the model as it does not imply a large amount of fine-tuning: the top
partner, which gives the most dangerous corrections to electroweak precision
observables,  can be relatively light as it can arise at a lower scale than $f$
by a suitable election of the model parameters.

We are  now ready to discuss the numerical results for the decays of the extra
neutral gauge boson.

\subsection{Littlest Higgs model}

\begin{figure}
 \centering
\includegraphics[width=3.5in]{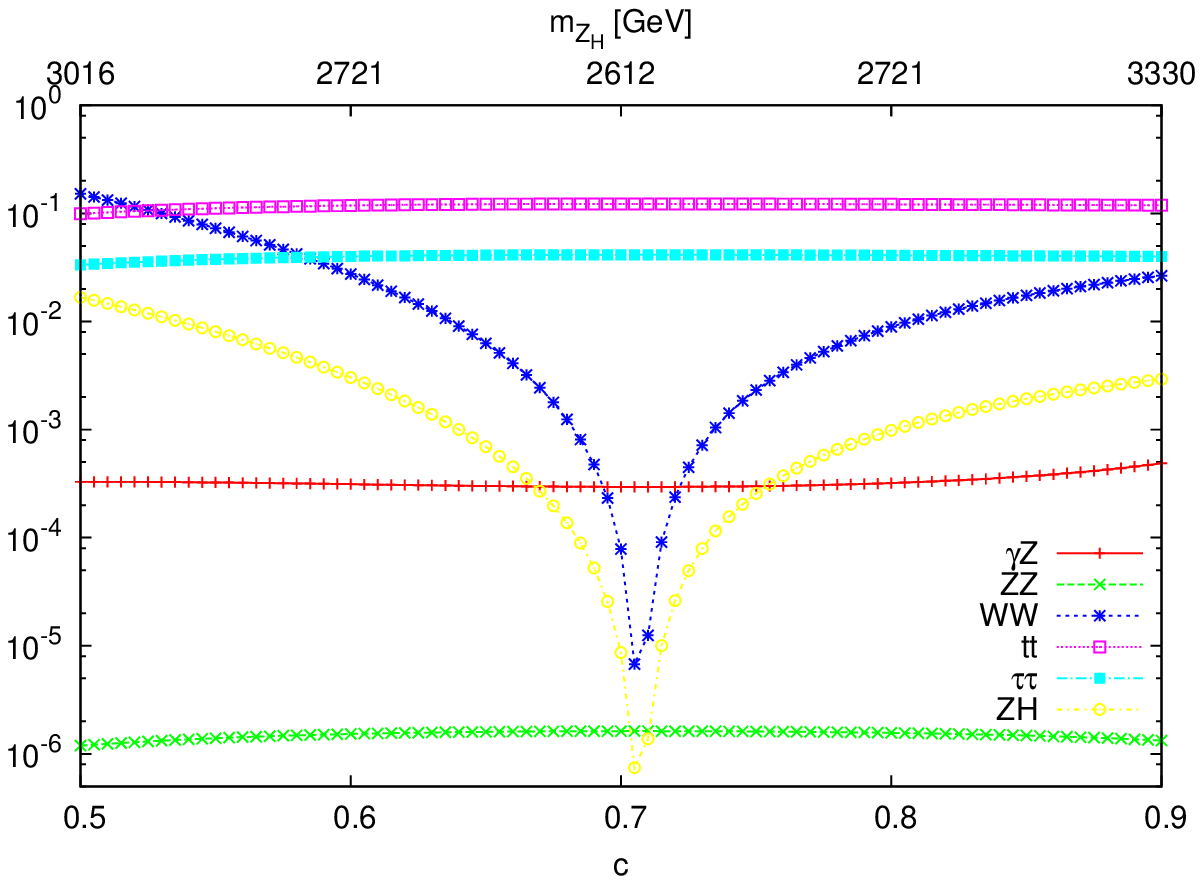}\includegraphics[width=3.5in]{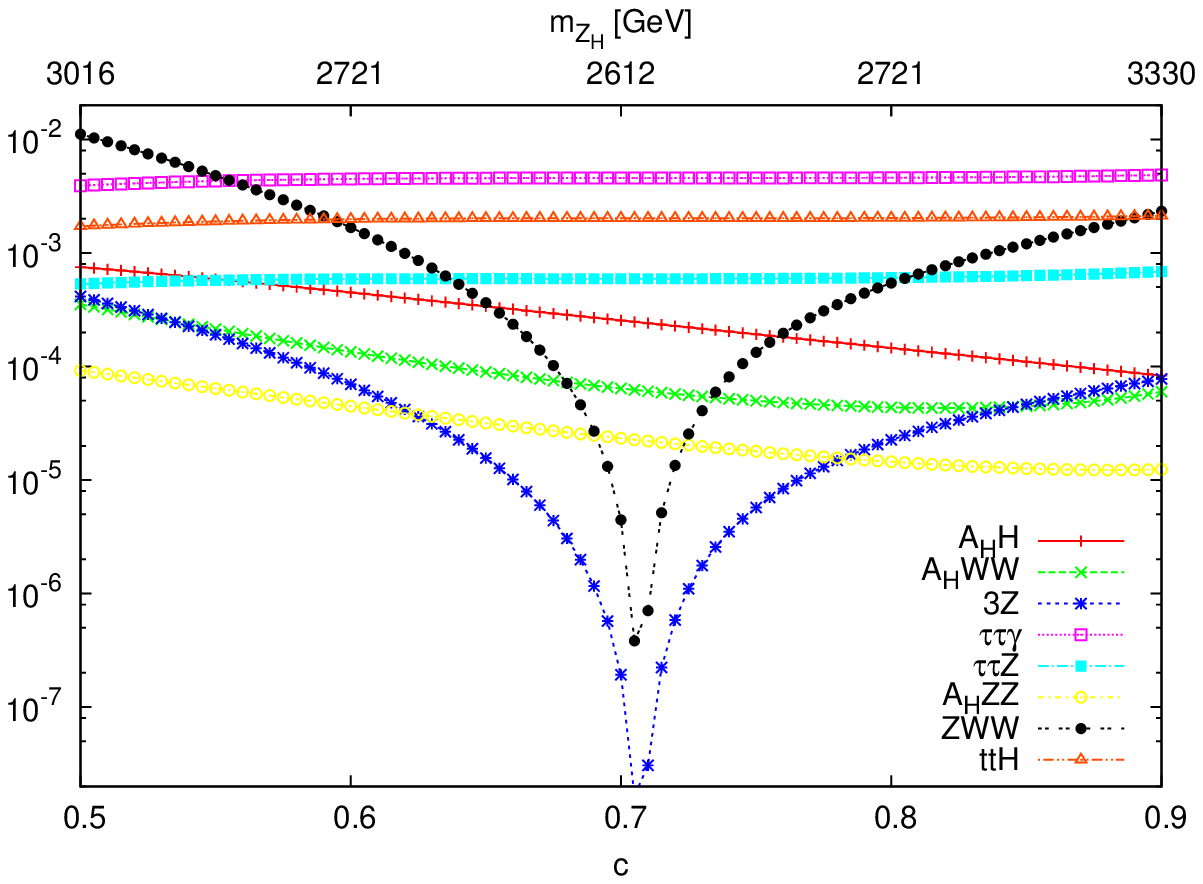}
 \caption{\label{brcLHM} Branching ratios for the one-loop decays $Z_H\to \gamma
Z$ and $Z_H\to ZZ$ in the LHM as a function of the mixing angle $c$. We also
include the main tree-level two- and three-body decays. We used the value
$m_H=120$ GeV for the Higgs boson mass. The branching ratios for the decays
$Z_H\to ZHH$, $Z_H\to \gamma WW$, and $Z_H\to A_H HH$
 are not shown in the plot but  $Br(Z_H\to ZHH)\sim 
Br(Z_H\to \gamma WW)\sim Br(Z_H\to ZWW)$ and $Br(Z_H\to A_HHH)\sim Br(Z_H\to
A_HZZ)$. For the one-loop decays we used the package LoopTools
\cite{'tHooft:1978xw,vanOldenborgh:1989wn,Hahn:2000jm} to numerically evaluate
the Passarino-Veltman scalar functions. 
}
 \end{figure}

In this model the $Z_H$ boson decay widths have a strong dependence on the
mixing angle of the gauge sector, namely, $\tan \theta=s/c$.  We will first
analyze the results for the $Z_H$ decays as functions of the mixing angle $c$
and for a particular value of $f$.  Then we  will examine the dependence of the
$Z_H$ branching ratios on the scale of the symmetry breaking $f$ for a fixed
value of $c$. In Fig. \ref{brcLHM}  we show the corresponding results for the 
$Z_H$ branching ratios discussed above as a function of the mixing angle $c$ and
for $f=4$ TeV, which corresponds to the strongest constraint on the scale $f$.
For simplicity we used $\tan\theta'=1$, although there is little dependence on
this parameter. Also, the value $120$ GeV is used for the mass of the Higgs
boson. We observe that around $c=1/\sqrt{2}$, the $Z_H$ gauge boson decays
mainly into a fermion pair. Due to the color number, the decay into a quark pair
is slightly dominant over the leptonic one. Apart from the color factor, the
branching ratios for the light fermions are almost identical and has little
dependence on the fermion mass. The subdominant decays are $Z_H\to \bar{t}tH$,
$Z_H\to\bar{f}f\gamma$, $\bar{f}{f}Z$, $Z_H\to A_H H$, $Z_H\to A_H WW$,  $Z_H\to
A_H ZZ$, and $Z_H\to A_H HH$. The branching ratios for the last two decays are
of similar size and the latter was not included  in the
plot. Other tree-level decays such as $Z_H\to WW$, $Z_H\to ZH$, $Z_H\to ZHH$,
$Z_H\to ZWW$, and $Z_H\to \gamma WW$ exactly vanish when $c=1/\sqrt{2}$. On the
other hand, when $c$ is far from $1/\sqrt{2}$, the decay  $Z_H\to WW$ width can
be as dominant  as the fermion decays. As far as the one-loop decays are
concerned, the $Z_H\to \gamma Z$ branching ratio can be as high as the
tree-level decays $Z_H\to A_H H$ or $Z_H\to \bar{ l } l  Z$, but the $Z_H\to
ZZ$ decay has a very small branching ratio. The former has a branching ratio of
the order of $10^{-3}$ while the latter has a rate of about $10^{-5}$.

\begin{figure}
 \centering
\includegraphics[width=3.5in]{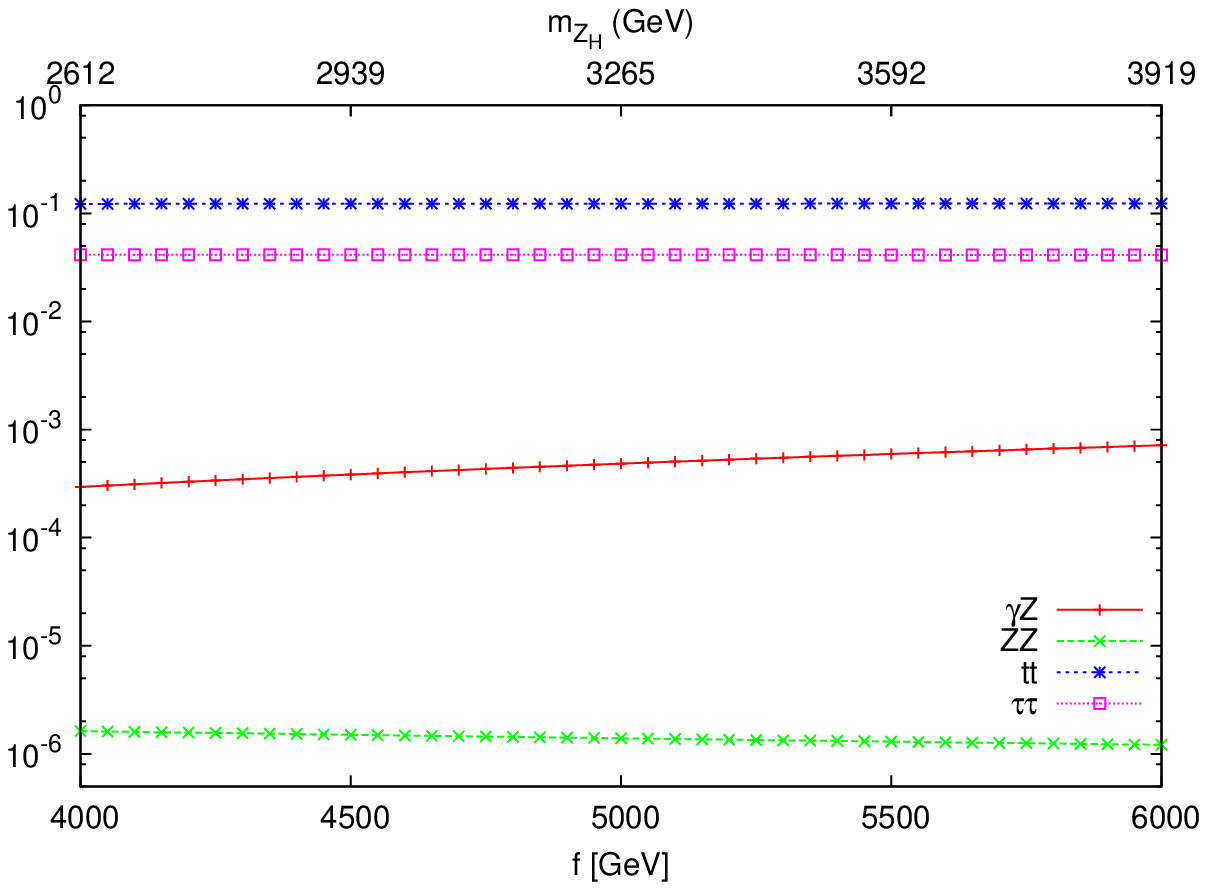}\includegraphics[width=3.5in]{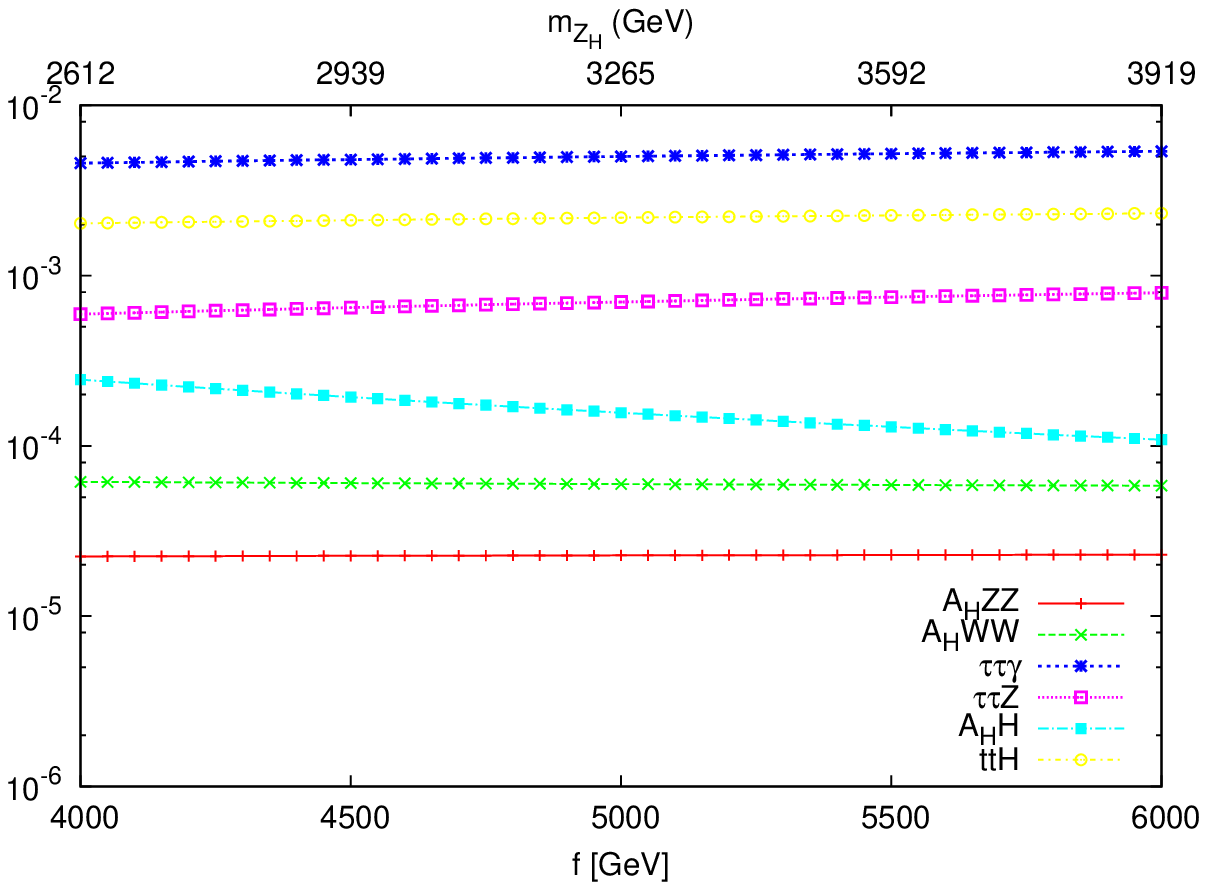}
 \caption{\label{brfLHM} Branching ratios for the one-loop decays $Z_H\to \gamma
Z$ and $Z_H\to ZZ$ in the LHM as a function of the scale of symmetry breaking
$f$ and for $c=1/\sqrt{2}$. We also include the main tree-level two- and
three-body decays. We used the value $m_H=120$ GeV for the Higgs boson mass.
The branching ratios not shown vanish exactly for this value of $c$.}
 \end{figure}

We now set the mixing angle at the value $c=1/\sqrt{2}$ and plot the branching
fractions for the $Z_H$ decays as functions of the scale $f$, as  shown in Fig.
\ref{brfLHM}. Several decay channels vanish in this scenario as commented above
due to the vanishing of the couplings involved in the decay amplitude. It is
interesting to note that the  $Z_H\to \gamma Z$  branching ratio is even larger
than the ones for the tree-level decays involving a heavy photon. The latter are
suppressed by phase space due to the large value of $m_{A_H}$. However, the
decay $Z_H\to ZZ$ has a negligible branching ratio and it would hardly have the
chance of being detected. 

\subsection{Simplest Little Higgs}
\begin{figure}
 \centering
\includegraphics[width=3.5in]{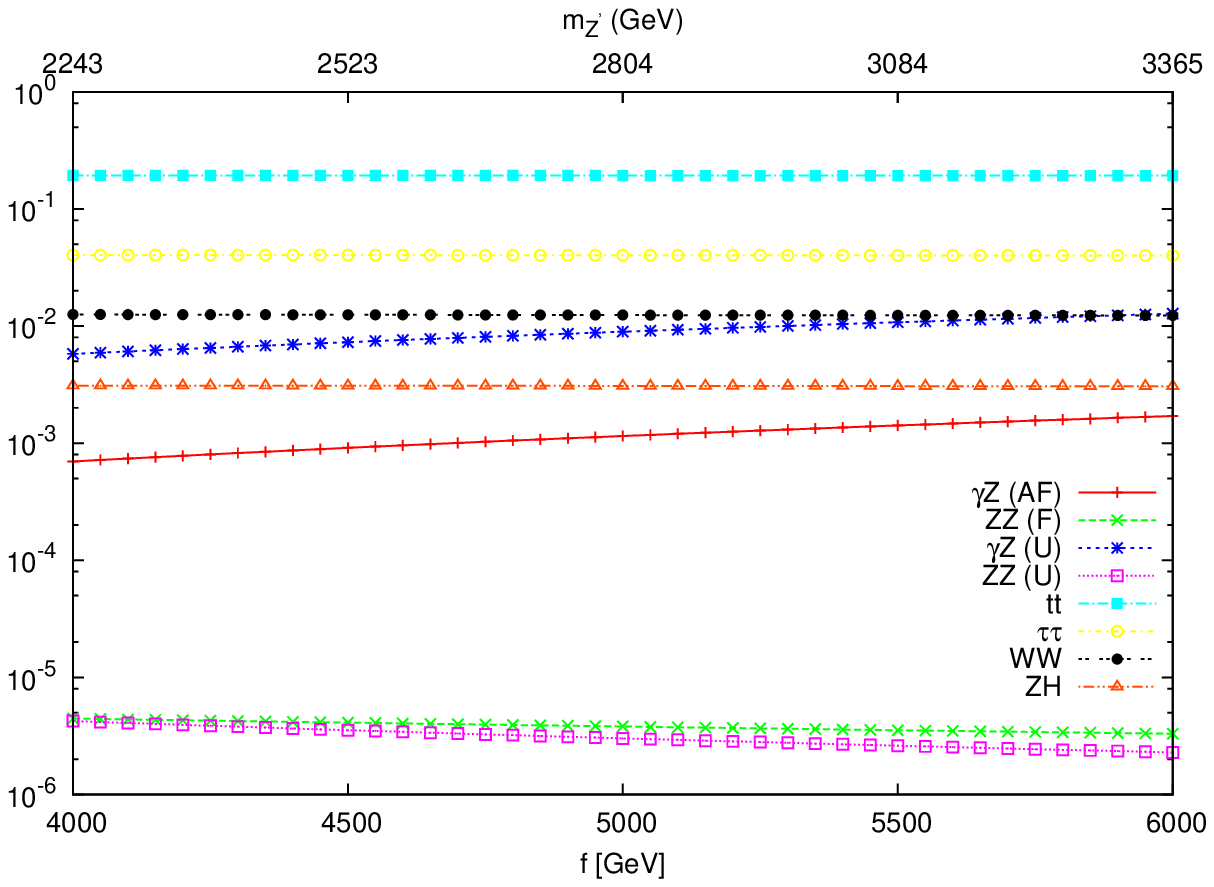}\includegraphics[width=3.5in]{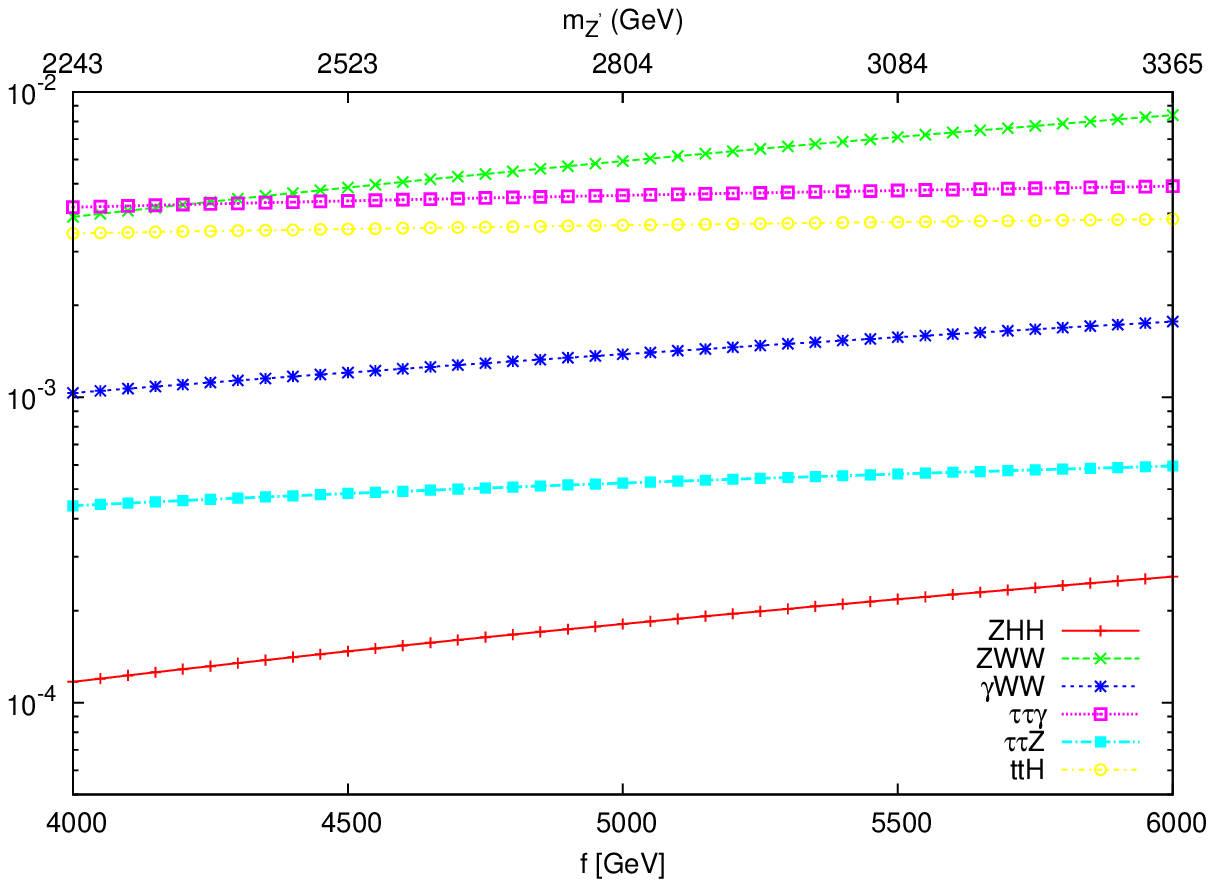}
 \caption{\label{brfSLH} Branching ratios for the one-loop decays $Z'\to \gamma
Z$ and $Z'\to ZZ$ in the SLHM as a function of the scale of symmetry breaking
$f$ in the universal (UE) and anomaly-free (AF) embeddings of the fermions. We
also include the main tree-level two- and three-body decays, which were obtained
using the couplings of the anomaly-free embedding. We used the value $m_H=120$
GeV for the Higgs boson mass.}
 \end{figure}

In this model, apart from the symmetry breaking scale $f$, the $Z'$ couplings
have no dependence on additional free parameters.
The branching ratios for the decays $Z'\to \gamma Z$ and $Z'\to ZZ$ are shown in
Fig. \ref{brfSLH} as functions of the symmetry breaking scale in the
anomaly-free embedding and also in the universal embedding. The branching ratios
for the main decays arising at the tree-level are also shown. We can observe
that the $Z'$ boson would decay mainly into a quark-antiquark pair, with a 
branching ratio larger than the one for the leptonic decays due to the factor
arising from the color number. The $Z'\to WW$
branching ratios is about one order of magnitude below and other decay channels
such as $Z'\to ZH$, , $Z'\to \bar{t}tH$ and $Z'\to
\bar{ l } l \gamma$,  $Z'\to ZHH$, $Z'\to \bar{ l } l  Z$, $Z'\to ZWW$, and
$Z'\to AWW$ have a smaller branching ratio, though they may be at the reach of
detection. All these decay channels are always present, in contrast with the
case of the LHM, where several decays are absent when $c=1/\sqrt{2}$. As far as
the one-loop decay $Z'\to ZZ$ are concerned, their
branching ratios are of the order of $10^{-5}$ in both the anomaly-free and the
universal embedding, while the decay $Z'\to \gamma Z$ has a rate of about
$10^{-3}$ in the anomaly-free embedding but is about one order of magnitude
larger in the universal embedding. While the $Z'\to
ZZ$ branching ratio remains almost unchanged in both the anomaly-free and the
universal embeddings, the $Z'\to\gamma Z$ branching ratio depends strongly on
the anomaly cancellation mechanism.

\subsection{Experimental perspectives}
Given the recent CDF, ATLAS, and CMS results, we expect conclusive news about the existence  of  a $Z^\prime$ gauge boson as long as its mass is between 1 and 3 TeV. LHC experiments, by using a total integrated luminosity of \rm 10~fb$^{-1}$, will be able to discover or rule out a neutral gauge boson  with a mass up to  2-3 TeV at 95\% C.L. \cite{Aaltonen:2011gp, Chatrchyan:2011wq, atlascolab}. Those experiments  have been  optimized to look for  signals from  narrow spin-1 resonances, like
our hypothetical heavy neutral gauge boson, decaying into electron or muon pairs. Recently, CDF \cite{Aaltonen:2011gp} , CMS \cite{Chatrchyan:2011wq}, and ATLAS\cite{atlascolab} have reported a lower limit on the  $Z^\prime$ mass of the order of 1 TeV  assuming SM-like $Z^\prime$ couplings to fermions. In such a scenario, the so-called sequential standard model (SSM), ATLAS and CMS have explored the potential discovery of a $Z^\prime$ decaying into leptons, for several $m_{Z^\prime}$ values,  at $\sqrt s= 7$, and $\sqrt{s}=14$ TeV \cite{Diener:2010sy, Petriello:2008pu}.

At the LHC, the  production of an extra neutral gauge boson would proceed mainly via the
Drell-Yan process \cite{Burdman:2002ns,Han:2003wu,Han:2005ru}. Using the branching ratios
obtained above in the LHM and the SLHM, we have calculated the number of $V\to V_i Z$
events at $\sqrt s=14$ TeV and an integrated luminosity of $\mathcal{L}= 100$ fb$^{-1}$.
For comparison purpose we also included the expected number of dilepton events. For the
model parameters, we have used the values $f=4$ TeV and $c= 1/\sqrt 2)$. These values
correspond to $m_{Z_H}=2.6$ TeV in the LHM and $m_{Z'}=2.2$ TeV in the SLHM. The results
are shown in Tables \ref{LHMtable}  and \ref{SLHMtable}.  Bearing in mind a comparison
between our calculations and the published ATLAS and CMS results, we  applied the average
correction factor $A\times \epsilon= 0.254$ value reported in \cite{Aad:2009wy} ($A$ is
the geometrical acceptance and $\epsilon$ is the reconstructed efficiency for the
$Z^\prime \to \bar l l$ channel) to the number of expected $V \to \bar l l$ events.

\begin{table}[htbp]
\centering
\begin{tabular}{|c|c|c|}
\hline
\multicolumn{3}{|c|}{LHM}
\\ \hline
\multirow{3}{*}{ }Decay channel  & $m_{Z_H}=2.612$ TeV &$N_{candidate}$ \\
\hline
\multirow{3}{*} {} $\sigma \times Br(Z_H \to l \bar l)$  & $20.5$ fb & $520$ 
\\
\hline
\multirow{3}{*} {} $\sigma \times Br(Z_H \to \gamma Z)$ & $0.147$ fb & $14$  \\
\hline
\multirow{3}{*} {} $\sigma \times Br(Z_H \to Z Z)$ & $8.12\times 10^{-4}$ fb &
$0$  \\
\hline
\end{tabular}
\caption{Expected $\sigma(pp\to Z_H) \times Br(Z_H \to\bar l l)$  and $\sigma (pp\to Z_H) \times Br(Z_H \to V_iZ)$ values and predicted number of candidate events for $m_{Z_H}=2.6$ TeV at $\sqrt s= 14$ TeV and $\mathcal{L}= 100$ fb$^{-1}$. For the $Z_H \to \bar l  l$ decay channel we included the experimental factor $A\times \epsilon= 0.254$ to obtain the event number  \cite{Aad:2009wy}.}
\label{LHMtable}
\end{table}

\begin{table}[htbp]
\centering
\begin{tabular}{|c|c|c|}
\hline
\multicolumn{3}{|c|}{Anomaly-free Embedding}
\\ \hline
\multirow{3}{*}{ }Decay channel  & $m_{Z'}=2.2$ TeV &$N_{candidate}$ \\
\hline
\multirow{3}{*} {} $\sigma \times Br(Z' \to l \bar l)$  & $4.04$ fb & $101$  \\
\hline
\multirow{3}{*} {} $\sigma \times Br(Z' \to \gamma Z)$  & $6.9 	\times
10^{-2}$ fb & $7$ \\
\hline
\multirow{3}{*} {} $\sigma \times Br(Z' \to Z Z)$  & $4.4\times 10^{-4}$ fb &
$0$ \\
\hline
\hline
\multicolumn{3}{|c|}{Universal Embedding } \\
\hline
\hline
\multirow{3}{*}{ }Decay channel  & $m_{Z'}=2.2$ TeV &$N_{candidate}$ \\
\hline
\multirow{7}{*} {} $\sigma \times Br(Z' \to \gamma Z)$  & $1.11 $ fb
& $115$ \\
\hline
\multirow{3}{*} {} $\sigma \times Br(Z' \to Z Z)$  & $4.23\times 10^{-4}$ fb &
$0$ \\
\hline
\end{tabular}
\caption{The same as in Table \ref{LHMtable} but for the extra neutral gauge boson of the SLHM. }
\label{SLHMtable}
\end{table}

 It is clear that there are promising expectations for the discovery of an extra
neutral gauge boson decaying into a lepton pair, in both the LHM and the SLHM. 
As soon as the LHC reaches the nominal  $\sqrt s$=  14 TeV energy, it will take
a few years to collect 100 fb$^{-1}$ of integrated luminosity. For $m_{Z_H}=2.6$
TeV and $m_{Z'}=2.4$ TeV,  which are the corresponding  values for the chosen
value of $f$, the expectations are of the order of a few hundred events in the
LHM and a few dozens in the SLHM.  These results  are similar to those obtained 
in other  models \cite{Nath:2010zj,Langacker:2009su,Rizzo:2006nw}, though they
are one or two orders of magnitude smaller than those reported by ATLAS and CMS,
using the  SSM model. As far as the potential observation of the  $ \gamma Z$
and $ZZ$ decay channels is concerned,  LHC experiments have studied  the
sensitivity to the production of SM diboson events, including  the $\gamma Z$
and $ZZ$ signals with the $Z$ gauge boson decaying into a highly  energetic 
lepton pair ($ee$ and $\mu\mu$) accompanied by an isolated high $p_T$ photon
\cite{Chatrchyan:2011rr, Aad:2011tc, Outschoorn:2010pt, Aaltonen:2010gy}.  The
$s$-channel $\gamma Z$ production is of special interest due to its sensitivity
to the $Z\gamma V_i$ ($V_i= Z, \gamma$) vertex \cite{Larios:2000ni},  which is
forbidden at tree-level in the SM.  From Table \ref{LHMtable}, we observe that
there would be about 14 $Z_H\to Z\gamma$ events, which gives some room for a
detailed data analysis. For  a heavier $Z_H$, an increase of about one order of
magnitude in the integrated luminosity  would be necessary to look for this
decay process. As for the  $Z_H \to Z Z$ decay channel, where the final state
includes four leptons, it is clear that an experimental signature is not
favorable. However, if we consider that the total LHC integrated luminosity will
be  of the order of $3000$ fb$^{-1}$, there is still some chance to observe  the
$Z_H \to Z Z$ decay channel. The results for the detection of these decays
channels in the framework of the SLHM  depend on the
 mechanism of anomaly cancellation, which can enhance considerably the
$Z'\to \gamma Z$ branching ratio. While the $Z'\to ZZ$ decays has
little chances of being detected, the detection of the $Z'\to\gamma Z$
decay would depend on the ``true`` mechanism of anomaly
cancellation. For instance, in the universal embedding there is more chances of
detecting the $Z'\to \gamma Z$ decay. Note that the cross section from
production of a $Z'$ gauge boson is also larger in the SLHM with universal
embedding as compared to the version with anomaly-free embedding.

In the case of $e^+e^-$ linear colliders \cite{Djouadi:2007ik}, we have calculated the $\sigma (e^- e^+ \to V_iZ)$ in both the LHM and the SLHM. For the  same $f$ and $c$ parameter values chosen above, a promising scenario for the discovery of the $ Z_H \to \gamma Z$ and $ Z_H \to ZZ$ decay channels  would be  an accelerator machine with  $\sqrt s= 2$ TeV and an
integrated luminosity of the order of 1000 $fb^{-1}$. With these conditions, we
obtain approximately 10 $ Z_H \to \gamma Z$ events   but less than one  $ Z_H
\to Z Z$ event. In
general, the possible detection of the $V\to V_i Z$ decays would improve largely
for a center-of-mass energy near the $V$ resonance. In this respect, the CLIC
potential is very promising for detecting several decays of an extra neutral
gauge boson with a mass of the order of $1-3$ TeVs.

\section{Conclusions and outlook}
We have calculated the one-loop decays $Z_H\to V_i Z$ ($V_i=\gamma, Z$) in the
framework of two popular versions of the little Higgs model. Since these decays
depend strongly on the mechanism of anomaly
cancellation, their study would provide complementary information that could help us 
to unravel the underlying theory. While the branching ratio for the $Z_H\to
\gamma Z$ decay can be as large as $10^{-3}$ in the LHM, the $Z_H\to ZZ$ decay
has a branching ratio of the order of $10^{-5}$ for $f=$ 4 TeV. In the SLHM
the
$Z'\to \gamma Z$ branching ratio is about $10^{-3}$ in the anomaly free
embedding but can be enhanced by about one order of magnitude in the universal
embedding. The decay $Z'\to ZZ$ is very suppressed and have branching ratios of
the order of $10^{-5}$ in both the anomaly-free and the universal embedding.
These class of decays are forbidden in the LHM with T-parity. However, the
$Z_H\to V_i A_H$ ($V_i=\gamma, Z$) decays can be of interest.
These decays proceed through triangle diagrams that include two different
fermions in the loop (one T-odd and one T-even) and the calculation is more
involved. The results for these decays will be presented elsewhere.

We  have also discussed the prospects for the experimental observation of the
$V\to V_i Z$ decays at the LHC. While the detection of an extra neutral gauge
boson, with a mass of about 3 TeV, decaying into a lepton pair looks very
promising in both the LHM and the SLHM, the situation for the $V \to V_i Z$
decays would be less favorable. In fact, it would be necessary to collect more
than 1000 fb$^{-1}$ of integrated luminosity in order to have few $Z_H\to \gamma
Z$ candidate events. The observation of the $Z_H\to ZZ$ decay would be even
less favorable. As far as the situation in the SLHM is concerned,  since the
$Z'\to
V_i Z$ decays are highly dependent on the mechanism of anomaly cancellation, a
more detailed knowledge of this mechanism would be necessary to asses the
possibility of detection of these decay modes. As far as the prospects at a
future $e^+e^-$ collider are concerned, it would be required a center-of-mass
energy near the $Z_H$ resonance to allow the detection of rare decays of an
extra neutral gauge boson, as the ones discussed in this work.

\acknowledgments{We acknowledge support from Conacyt and SNI (M\'exico). Support from VIEP-BUAP is also acknowledged.}

\appendix

\section{Couplings of the extra neutral gauge boson in little Higgs models}
\label{Couplings}
In this appendix we collect all the Feynman rules necessary for our calculation in  the
unitary gauge. They were taken from Refs. \cite{Burdman:2002ns,Han:2003wu} and
\cite{Han:2005ru}.

\subsection{Couplings to the light and heavy fermions}

The coupling of the extra neutral gauge boson $V$ to a fermion pair can  be written as
\begin{equation}
{\cal L}=  -\frac{ig}{c_W }\bar{f_i}\gamma^{\mu}\left({g'}^f_L P_L+{g'}^f_R P_R\right)f_j{V}_\mu,
\label{ZHffLag}
\end{equation}
with $P_{L,R}$ the usual chirality projectors.

In the LHM, the couplings of the $Z_H$ gauge boson to SM fermions are universal and are given by ${g'}_{L}^f=\frac{c_W\,c}{s} T^3$ and ${g'}_{R}^f=0$, where $T^3_f = 1$ $(-1)$ for up (down) type fermions. Although the $Z_H$ gauge boson couples to a top quark partner pair, this coupling is very suppressed as it arises  up the order of $(v/f)^2$. The same is true for  the nondiagonal coupling  $Z_H \bar{t}T$, which  is of the order of $v/f$. We have neglected those Feynman diagrams mediated by those couplings and have only considered couplings to fermions of the same flavor.

As far as the SLHM is concerned, the couplings of the $Z'$ gauge boson to the fermions depend on the embedding of  the heavy fermions. In Table \ref{SLHMZff} we collect the $Z'$ couplings to SM fermions in both the universal and anomaly-free embeddings, whereas the couplings to heavy fermions are shown in Table \ref{SLHMZffh}.  On the other hand, the $Z$ gauge boson couplings to SM fermions are the same as the SM ones but corrected by terms of the order of $(v/f)^2$. The couplings of the $Z$ boson to heavy fermions are also presented in  Table \ref{SLHMZffh}.

\begin{table}[!h]
\centering
\begin{tabular}{|c|c|c|c|c|}
\hline
&\multicolumn{2}{|c|}{Universal}&\multicolumn{2}{|c|}{Anomaly-free embedding}\\
\cline{2-5}
\hline
	& $ f^f_L$&$ f^f_R$&$ f_L^f$&$f_R^f$\\
\hline
$Z^{\prime} \overline \nu_i \nu_i$ & $\frac{1}{2} - s_W^2$&0&$\frac{1}{2} - s_W^2$&0\\
\hline
$Z^{\prime} \overline e_i e_i$ & $\frac{1}{2} - s_W^2$ &$- s_W^2$ &$\frac{1}{2} - s_W^2$&$- s_W^2$\\
\hline
$Z^{\prime} \overline u_i u_i$ ($i=1,2$)&$\frac{1}{2} - \frac{1}{3}s_W^2$ &$ \frac{2}{3}s_W^2$
& $-\frac{1}{2} + \frac{2}{3}s_W^2$ &$ \frac{2}{3}s_W^2$   \\
\hline
$Z^{\prime} \overline d_i d_i$ ($i=1,2$)& $\frac{1}{2} - \frac{1}{3}s_W^2$ &$- \frac{1}{3}s_W^2$
& $-\frac{1}{2} + \frac{2}{3}s_W^2$ &$- \frac{1}{3}s_W^2$  \\
\hline
$Z^{\prime} \overline t t$ & $\frac{1}{2} - \frac{1}{3}s_W^2$ &$ \frac{2}{3}s_W^2$ & $\frac{1}{2} - \frac{1}{3}s_W^2$ &$ \frac{2}{3}s_W^2$\\
\hline
$Z^{\prime} \overline b b$ & $\frac{1}{2} - \frac{1}{3}s_W^2$ &$- \frac{1}{3}s_W^2$& $\frac{1}{2} - \frac{1}{3}s_W^2$ &$- \frac{1}{3}s_W^2$ \\
\hline
\end{tabular}
\caption{Couplings of the $Z'$ gauge boson to
the SM fermions in the  universal and anomaly-free embeddings of the SLHM. Unless stated otherwise $i=1..3$. The constants appearing in Eq. (\ref{ZHffLag}) are given by ${g'}_{L,R}^f=\frac{f_{L,R}^f}{\sqrt{3-4s_W^2}}$. \label{SLHMZff}}
\end{table}

\begin{table}[!h]
\centering
\begin{tabular}{|c|c|c|c|}
\hline
&\multicolumn{2}{|c|}{$Z'$ couplings}&$Z$ couplings\\
\cline{2-4}
\hline
 & ${f_L}^f$ &${f_R}^f$&$g_L^f=g_R^f$   \\
\hline
$V \bar{N}_i N_i$  & $-1 + s_W^2$ &$0$&0\\
\hline
$V \bar{T} T$ & $-1 + \frac{5}{3}s_W^2$& $\frac{2}{3}s_W^2$ &$-\frac{2}{3}s_W^2$\\
\hline
$V \bar{U} U$, $V \bar{C} C$ &$-1 + \frac{5}{3}s_W^2$ &$
\frac{2}{3}s_W^2$&$-\frac{2}{3}s_W^2$\\
\hline
$V \bar{D} D$, $V \bar{S} S$& $1 - \frac{4}{3}s_W^2$&$ -
\frac{1}{3}s_W^2$&$\frac{1}{3}s_W^2$\\
\hline
\end{tabular}
\caption{Couplings of the $Z'$ and $Z$ gauge bosons to the fermion partners in the SLHM. The next-to-last line is for the universal embedding and the last line is for the anomaly-free embedding.  The constants appearing in Eq. (\ref{ZHffLag}) are given by ${g'}_{L,R}^f=\frac{f_{L,R}^f}{\sqrt{3-4s_W^2}}$. \label{SLHMZffh}}
\end{table}

\subsection{Couplings to SM gauge bosons and the Higgs boson}

We now present the couplings of the extra neutral gauge boson to the Higgs boson and the SM gauge bosons.

The $V_1^\mu V_2^\nu H$ and $V_1^\mu V_2^\nu HH$ couplings, with $V_{1,2}$ standing for a neutral gauge boson, can be written as

\begin{eqnarray}
V_1^\mu V_2^\nu H&=& i g_{V_1 V_2 H} g^{\mu\nu}\\
\label{V1V2H}
V_1^\mu V_2^\nu HH&=&i g_{V_1 V_2 HH} g^{\mu\nu}.
\label{V1V2HH}
\end{eqnarray}
The Feynman rule for the trilinear gauge boson vertex, with all particles outgoing, is given by

\begin{equation}
V_1^{\mu}(k_1) V_2^{\nu}(k_2) V_3^{\rho}(k_3)= i g_{V_1V_2V_3} \left( g^{\mu\nu} (k_1 - k_2)^{\rho}+ g^{\nu\rho} (k_2 - k_3)^{\mu} + g^{\rho\mu} (k_3 - k_1)^{\nu}\right),
\label{trilinear}
\end{equation}
whereas the quartic gauge boson coupling can be written as
\begin{equation}
 V_1^{\mu} V_2^{\nu} V_3^{\rho} V_4^{\sigma}= i g_{V_1V_2V_3} \left(2g^{\mu\nu}g^{\rho\sigma}-g^{\mu \rho}g^{\nu\sigma}-g^{\mu\sigma}g^{\nu\rho}\right).
\label{quartic}
\end{equation}
The corresponding couplings constants in both the LHM and the SLHM are shown in Table \ref{trilandquar}.

\begin{table}[!h]
\centering
\begin{tabular}{|c|c|c|}
\hline
 & LHM ($V=Z_H$) & SLHM ($V=Z'$) \\
 \hline
 $g_{VZH}$&$-\frac{g^2 v}{2c_W}\frac{g^2 (c^2-s^2)}{2\,c\,s}$&$\frac{g^2 v(1-t_W^2)}{4 c_W \sqrt{3-t_W^2}}$\\
\hline
$g_{VZHH}$& $-\frac{g^2 (c^2-s^2)}{4\,c\,s\, c_W}$& $\frac{g^2(1-t_W^2)}{4 c_W \sqrt{3-t_W^2}}$ \\
\hline
$g_{VWW}$& $ \frac{g  c\,s\,(c^2-s^2) \, v^2}{ 2  f^2}$& $ -\frac{g  (1-t_W^2)\sqrt{3-t_W^2} \, v^2}{ 8 f^2}$ \\
\hline
 $g_{WWVZ}$& $-\frac{g^2 (c_W^2 - s_W^2) sc (c^2 - s^2)v^2}{2c_W\,f^2}$&0\\
\hline
 $g_{WWVA}$& $-\frac{g^2 s_W sc (c^2 - s^2)v^2}{f^2}$&0\\
\hline
\end{tabular}
\caption{Trilinear and quartic couplings of the neutral gauge boson $V$ in little Higgs models. \label{trilandquar}}
\end{table}
\subsection{Couplings involving other heavy gauge bosons}
We first present the couplings involving the heavy photon and the heavy charged gauge boson arising in the LHM. These couplings are necessary for the calculation of various tree-level three-body decays of the extra neutral gauge boson.  The Feynman rules for this kind of couplings are given in  Eq. (\ref{V1V2H}) through Eq. (\ref{quartic}).  We present the coupling constants for the LHM in Table \ref{AHZHcoup}. Notice that in the framework of the SLHM, the heavy photon is absent and there is no quartic couplings involving the heavy charged gauge boson $X$ with the extra neutral gauge boson $Z'$. Besides, the couplings $Z'XW$ and $ZXW$ vanish.

\begin{table}[!h]
\centering
\begin{tabular}{|c|c|}
\hline
 & LHM  \\
\hline
 $g_{ WWZ_H A_H}$& $-g^2x_H \frac{v^2}{f^2} $\\
\hline
$g_{Z_H A_HH}$&$-\frac{g g' v (s^2c^{'2}+c^2s^{'2})}{4\,c\,s\,c'\,s'}$\\
\hline
$g_{Z_H A_H HH}$& $-\frac{g g' (s^2c^{'2}+c^2s^{'2})}{4\,c\,s\,c'\,s'}$\\
\hline
$g_{Z_HW_HW}$&$g$\\
\hline
$g_{ZW_HW}$&$g x_W^Z\frac{v^2}{f^2}$\\
\hline
\end{tabular}
\caption{Couplings involving the heavy gauge bosons $A_H$ and $W_H$ along with SM bosons in the LHM.  \label{AHZHcoup}}
\end{table}

Finally, we present some additional couplings necessary for the calculation of the $Z_H$ decays in Table \ref{AHcoup}. Some SM couplings involved in the calculation of the decays of the extra neutral gauge boson receive corrections of the order of $(v/f)^2$  but they were neglected from our calculation.
\begin{table}[!h]
\centering
\begin{tabular}{|c|c|}
\hline
 & LHM  \\
\hline
 $g_{A_H WW}$& $g c_W x_Z^B \frac{v^2}{f^2} $\\
\hline
$g_{A_H A_H H}$& $-\frac{g' v^2}{2}$\\
\hline
$g_{A_H W_H W}$& $gx_H\frac{v^2}{f^2}$\\
\hline
\end{tabular}
\caption{Couplings involving the heavy photon $A_H$ and SM particles. Here $x_Z^B=-\frac{5}{2s_W}s'c'(c^{'2}-s^{'2})$ and $x_H=\frac{5}{2}g g'\frac{sc s'c'(s^2c^{'2}+c^2s^{'2})}
{5g^2s^{'2}c^{'2}-g^{'2}s^2c^2}$. \label{AHcoup}}
\end{table}

\end{document}